\documentclass[11pt]{article}
\usepackage{amsfonts}
\usepackage{amssymb}
\newsymbol\contra 1379
\def\a{\alpha}
\def\b{\beta}
\def\g{\gamma}

\def\o{\omega}
\def\p{\partial}

\def\s{\sigma}
\def\th{\theta}

\def\hra{\hookrightarrow}

\def\ca{\mathcal A}
\def\cb{\mathcal B}
\def\cd{\mathcal D}

\def\cg{\mathcal G}
\def\ch{\mathcal H}

\def\cm{\mathcal M}
\def\cn{\mathcal N}

\def\ct{\mathcal T}

\def\cv{\mathcal V}
\def\C{\mathbb C}

\def\R{\mathbb R}

\def\Z{\mathbb Z}
\newcommand{\Ad}{{\mathop{\mbox{ad}}\nolimits}}
\newcommand{\ad}{{\mathop{\mbox{Ad}}\nolimits}}
\newcommand{\Int}{{\mathop{\mbox{Int}}\nolimits}}
\newcommand{\Ker}{{\mathop{\mbox{Ker}}\nolimits\,}}

\newcommand{\Tr}{{\mathop{\mbox{Tr}}\nolimits\,}}
\newcommand{\fm}{{\mathfrak M}}

\newcommand{\fg}{{\mathfrak g}}

\begin{document}
\begin{flushright}
 hep-th/9909135
\end{flushright}

\vskip 1.5cm

\begin{center}
{\LARGE \bf Holomorphic Analogs of }
\vskip 0.5cm
{\LARGE \bf Topological Gauge Theories$^*$}
\vskip 1.5cm
{\Large Alexander D. Popov$^{1}$}

\vskip 1cm

{\em Bogoliubov Laboratory of Theoretical Physics\\
JINR, 141980 Dubna, Moscow Region, Russia}\\

\vskip 1.1cm
\end{center}

\begin{center}
{\bf Abstract}
\end{center}

\begin{quote}
We introduce a new class of gauge field theories in any complex
dimension, based on algebra-valued $(p,q)$-forms on complex
$n$-manifolds. These theories are holomorphic analogs of the
well-known Chern-Simons and BF topological theories defined on
real manifolds. We introduce actions for different special
holomorphic BF theories on complex, K\"{a}hler and Calabi-Yau
manifolds and describe their gauge symmetries. Candidate observables,
topological invariants and relations to integrable models are briefly
discussed.

\vskip 5mm

{
}
\end{quote}

\vfill
\hrule width 5.cm

\vskip 3mm

{\small \noindent ${}^{*\ {}}$
supported in part by the grant RFBR-99-01-01076

\vskip 1mm

\noindent ${}^{1\ {}}$
E-mail: popov@thsun1.jinr.ru}

\newpage

\section{Introduction}

Topological field theories~\cite{W1,W2} were intensively studied,
developed and generalized over the last ten years (see
e.g.~\cite{BBRT} and references therein). Among these theories,
of particular interest are non-Abelian Chern-Simons gauge
theories~\cite{W2}. The Chern-Simons (CS) action functional is $$
S_{\mathrm{CS}}=\int_X\Tr(AdA+\frac{2}{3}A\wedge A\wedge A),
\eqno(1.1) $$ where $X$ is an oriented smooth 3-manifold, $A$ is a
connection 1-form on a principal $G$-bundle $P$ over $X$, and $G$ is
a Lie group. The Euler-Lagrange equations for the action (1.1) are $$
F_A=0, \eqno(1.2) $$ where $F_A=dA+A\wedge A$ is the curvature of a
connection $A$. Thus, non-Abelian CS theories give a field-theoretic
description of flat connections on $G$-bundles over 3-dimensional
manifolds $X$, and expectation values of quantum observables for
these theories are topological invariants of $X$~\cite{W2,BBRT}.

\smallskip

Other interesting topological field theories associated with flat
connections are so-called BF theories~\cite{H} generalizing
Abelian models introduced in~\cite{S}. These theories can also be
considered as a generalization of Chern-Simons theories to arbitrary
dimensions. The classical action for a non-Abelian BF theory has the
following form:  $$ S_{\mathrm {BF}}=\int_X\Tr(B\wedge F_A),
\eqno(1.3) $$ where $X$ is an oriented smooth $n$-manifold, $F_A$ is
the curvature of a connection 1-form $A$ on a principal $G$-bundle
$P$ over $X$, and $B$ is an $(n-2)$-form with values in the adjoint
bundle $\Ad P=P\times_G \fg$. We consider a semisimple Lie group $G$
and denote by $\fg$ its Lie algebra. The variation of the action
(1.3) w.r.t. $B$ gives Eqs.(1.2), and the variation of the action
w.r.t. $A$ gives the equations $$ d_AB=0, \eqno(1.4) $$ where
$d_A=d+\Ad A$ is the covariant differential on $P$. Thus, BF theories
describe flat connections $A$ on bundles $P$ over $n$-manifolds $X$
and $\Ad P$-valued $d_A$-closed $(n-2)$-forms $B$.

\smallskip

An interesting class of (topological) gauge models has been introduced by
considering the classical limit of the string field theory for open
$N=2$ topological strings on Calabi-Yau 3-folds~\cite{W3, BCOV}.
The corresponding action~\cite{W3} is given by $$
S_{\mathrm{hCSW}}=\int_Z\th\wedge \Tr (A^{0,1}\bar\partial A^{0,1}
+\frac{2}{3} A^{0,1}\wedge A^{0,1}\wedge A^{0,1}), \eqno(1.5a) $$
where $\th$ is a nowhere vanishing holomorphic $(3,0)$-form on a
Calabi-Yau 3-fold $Z$, and $A^{0,1}$ is the $(0,1)$-component of a
connection 1-form $A$ on a principal $G$-bundle $P$ over $Z$. The
field equations following from the Chern-Simons-Witten (CSW) action
(1.5a) are $$ F^{0,2}_A=\bar\partial A^{0,1}+ A^{0,1}\wedge
A^{0,1}=0, \eqno(1.5b) $$ where $\bar\partial$ is the $(0,1)$-part of
the exterior derivative $d=\partial +\bar\partial$, and $F^{0,2}_A$
is the $(0,2)$-part of the curvature $F_A$ of a $G$-bundle $P$ with a
complex structure group $G$. Recently holomorphic CSW theories (1.5)
were discussed~\cite{DT,Th} in the frames of the program on extending
the results of Casson, Floer and Jones to Calabi-Yau (CY) threefolds.

\smallskip

Our discussion can be summarized in the following diagram:
$$
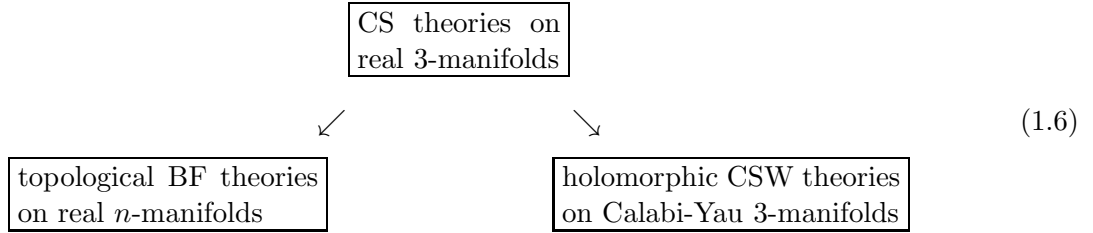
\fbox{\parbox{2.7cm}{CS theories on real
3-manifolds}}
$$
$$
\swarrow\hspace{3cm}\searrow
\eqno(1.6)
$$
$$
\fbox{\parbox{4cm}{topological BF theories on real
$n$-manifolds}} \hspace{3cm}
\fbox{\parbox{4.5cm}{holomorphic CSW theories on
Calabi-Yau 3-manifolds}}
$$
\medskip

\noindent
The arrows mean that one theory is a generalization of another one.
Notice that holomorphic CSW theories (1.5) are defined only in three
complex dimensions by analogy with the fact that CS theories (1.1)
with the action  linear in derivatives are defined only in three real
dimensions. At the same time, it is desirable to have a field theory
description of flat $(0,1)$-connections $A^{0,1}$ (holomorphic
structures) in arbitrary dimensions. The purpose of our paper is to
introduce {\it holomorphic BF theories} which generalize both topological
BF theories (1.3) and holomorphic CSW theories (1.5) to complex
$n$-manifolds and describe holomorphic structures on bundles in any
complex dimension.  We also introduce three {\it special} holomorphic
BF theories and describe, in particular, stable holomorphic bundles.
So, our goal in this paper is to extend the diagram (1.6) to the
diagram $$ \fbox{\parbox{2.7cm}{CS theories on real 3-manifolds}} 
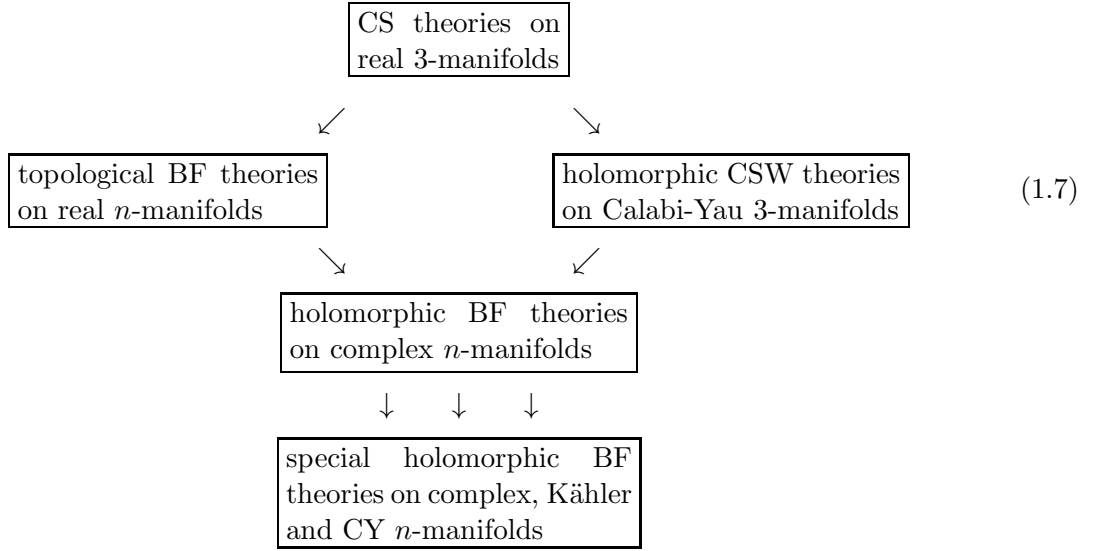
$$
$$ \swarrow\hspace{3cm}\searrow $$ $$ \fbox{\parbox{4cm}{topological
BF theories on real $n$-manifolds}} \hspace{3cm}
\fbox{\parbox{4.5cm}{holomorphic CSW theories on Calabi-Yau
3-manifolds}} \eqno(1.7) $$ $$ \searrow\hspace{3cm}\swarrow $$ $$
\fbox{\parbox{4.5cm}{holomorphic BF theories on complex
$n$-manifolds}} $$ $$ \downarrow\qquad  \downarrow\qquad \downarrow
$$ $$ \fbox{\parbox{4.6cm}{special holomorphic BF theories on
complex, K\"ahler and CY $n$-manifolds}} $$ \medskip

\noindent
which includes  holomorphic analogs not only of Chern-Simons but also of
BF theories.

\section{Holomorphic BF theories}
\subsection{Non-Abelian models}

Let $Z$ be a complex $n$-manifold, $G$ a (complex) semisimple Lie
group, $\fg$ its Lie algebra, $P$ a principal $G$-bundle over $Z$ and
$A^{0,1}$ the $(0,1)$-component of a connection 1-form $A$ on $P$.
Components $A^{0,1}$ of connections $A$ will be called $(0,1)$-{\it
connections}. The curvature $F_A$ of a connection $A$ splits into
components, $$ F_A=F^{2,0}_A+ F^{1,1}_A + F^{0,2}_A, $$ where
$F^{0,2}_A=\bar\partial A^{0,1}+A^{0,1}\wedge A^{0,1}$ is
the $(0,2)$-component of the curvature tensor and
$\bar\partial$ is the $(0,1)$-part of the exterior derivative
$d=\partial +\bar\partial$. We also split $d_A=d+\Ad A=\p_A+\bar\p_A$.

\smallskip

Let us consider the bundle of groups $\Int P=P\times_GG$ ($G$ acts on
itself by internal automorphisms: $h\mapsto ghg^{-1}$, $h,g\in G$)
associated with $P$.  We define the infinite-dimensional group of
gauge transformations as the group $\cg =\Gamma (Z, \Int P)$ of
global smooth sections of the bundle $\Int P$. We also consider the
adjoint bundle $\Ad P=P\times_G\fg$ of Lie algebras and denote by $
\Omega^{p,q}(Z,\Ad P) $ the space of smooth
$(p,q)$-forms on $Z$ with values in the bundle $\Ad P$,
$p,q,...=1,...,n=dim_\C Z$.

\smallskip

{}For simplicity we shall consider bundles $P$ that are equivalent to
the trivial one as  smooth bundles (i.e. topologically trivial),
$P\simeq Z\times G$. Therefore, we have $\Int P\simeq Z\times G$,
$\Ad P\simeq Z\times \fg$, and $\Ad P$-valued $(p,q)$-forms may be
considered as $\fg$-valued $(p,q)$-forms on $Z$. The space of such
forms is denoted by $ \Omega^{p,q}(Z,\fg)$.
Generalization to the case of topologically nontrivial bundles is
straightforward and not difficult.

\smallskip

Let us consider the $(0,2)$-component $F^{0,2}_A$ of the curvature
tensor of a connection 1-form on $P$ and a $\fg$-valued
$(n,n-2)$-form $B$ on $Z$. We introduce the action $$
S_{\mathrm{hBF}}=\int_Z \Tr(B\wedge  F^{0,2}_A), \eqno(2.1) $$ and
models with such actions will be called {\it holomorphic BF
theories}.  The field equations obtained from the action (2.1) are $$
\bar\partial A^{0,1}+A^{0,1}\wedge A^{0,1}=0, \eqno(2.2a) $$ $$
\bar\partial B+A^{0,1}\wedge B- B\wedge A^{0,1}=0.  \eqno(2.2b) $$
Solutions $A^{0,1}$ of Eqs.(2.2a) will be called {\it flat}
$(0,1)$-connections on $P$. Notice that Eqs.(2.2a) coincide with the
compatibility conditions $\bar\partial^2_A=0$ of Eqs.(2.2b). So,
holomorphic BF theories describe flat $(0,1)$-connections on $P$
which correspond to holomorphic structures on the bundle $P\to Z$.

\smallskip

By expanding $A^{0,1}$ and $B$ in the generators $\{J^i\}$ of the group
$G$, one can rewrite Eqs.(2.2) in the form $$ \bar\partial A^i
+\frac{1}{2}f^i_{jk} A^j\wedge A^k =0, \eqno(2.3a) $$ $$ \bar\partial
B^i +f^i_{jk} A^j\wedge B^k =0, \eqno(2.3b) $$ where $A^{0,1}=A^iJ_i,
B=B^iJ_i$, and $f^i_{jk}$ are structure constants of the group $G$,
$i,j,...=1,...,dim_\C G$. Notice that Eqs.(2.3b) can be rewritten in
the form $\bar\partial_A B =0$, where $\bar\partial_A=\bar\partial
+\Ad A^{0,1}$.

\smallskip

The gauge group $\cg$ acts on a $(0,1)$-connection $A^{0,1}$ on $P$
and on a field $B\in \Omega^{n,n-2}(Z,\fg )$ by the formulas $$
A^{0,1}\mapsto \ad_{g^{-1}}A^{0,1}= g^{-1}A^{0,1}g
+g^{-1}\bar\partial g, \eqno(2.4a) $$ $$ B\mapsto \ad_{g^{-1}}B=
g^{-1}Bg, \eqno(2.4b) $$ where $g\in\cg$. The action (2.1) is
invariant under the transformations (2.4) and under the following
``cohomological" symmetry transformation:  $$ B\mapsto B+
\bar\partial_A\Phi , \eqno(2.5) $$ where
$\Phi\in\Omega^{n,n-3}(Z,\fg)$. By virtue of this invariance,
$\bar\partial_A\Phi$ is a trivial solution of Eqs.(2.2b)
for any $\Phi\in\Omega^{n,n-3}(Z,\fg)$, and solutions $B$ and $B+
\bar\partial_A\Phi$ of Eqs.(2.2b) are considered as equivalent.

\smallskip

We denote by $\cn$ the space of solutions to Eqs.(2.2a) and by $\cm$
the set of orbits of the gauge group $\cg$ in the set $\cn$,
$$\cm =\cn /\cg .
\eqno(2.6a)
$$
Let $\pi$ be a projection
$$\pi :\cn\to\cm .
\eqno(2.6b)
$$
By definition, $\cn$ is the space of flat $(0,1)$-connections
(holomorphic structures) on the bundle $P$, and $\cm$ is the moduli
space of flat $(0,1)$-connections. Put another way, the moduli space
$\cm$ of holomorphic structures on $P$ is the space of gauge
inequivalent solutions to Eqs.(2.2a).

\smallskip

Equations (2.2b) are linear in $B$. For any fixed flat
$(0,1)$-connection $A^{0,1}$ the space $\cb_A$ of nontrivial
solutions to Eqs.(2.2b) is the $(n,n-2)$th Dolbeault cohomology group
$$
\cb_A=H^{n,n-2}_{\bar\partial_A;P}(Z):=\frac{\{B\in\Omega^{n,n-2}(Z,\fg ):
\bar\partial_A B=0\}}{\{B=\bar\partial_A\Phi , \Phi\in \Omega^{n,n-3}
(Z,\fg )\}},
\eqno(2.7)
$$
which is defined to be the space of equivalence classes of
$\bar\partial_A$-closed $\fg$-valued $(n,n-2)$-forms on $Z$, modulo
$\bar\partial_A$-exact $\fg$-valued $(n,n-2)$-forms. So, the space of
nontrivial solutions to Eqs.(2.2b) forms the {\it vector space}
$\cb_A$ depending on a solution $A^{0,1}$ of Eqs.(2.2a). Therefore
the space of solutions to Eqs.(2.2) forms a vector bundle
$\ct\to\cn$, the base space of which is the space $\cn$ of solutions
to Eqs.(2.2a), and fibers of the bundle $\ct$ at the points
$A^{0,1}\in\cn$ are the vector spaces
$H^{n,n-2}_{\bar\partial_A;P}(Z)=\cb_A$ of nontrivial solutions to
Eqs.(2.2b).

\smallskip

Recall that the gauge group $\cg$ acts on solutions $(A^{0,1},B)$ of
Eqs.(2.2) by formulas (2.4). Therefore, identifying points
$(A^{0,1},B)\in\ct$ and $(g^{-1}A^{0,1}g+g^{-1}\bar\partial g,
g^{-1}Bg)\in\ct$ for any $g\in\cg$, we obtain the moduli space $$ \fm
=\ct /\cg \eqno(2.8) $$ of solutions to Eqs.(2.2). The space $\fm$
is a vector bundle over the moduli space $\cm$ of flat
$(0,1)$-connections. If we denote by $[A^{0,1}]$ the gauge
equivalence class of a flat $(0,1)$-connection $A^{0,1}$, then the
fiber of this bundle at the point $[A^{0,1}]\in \cm$ will be
isomorphic to the Dolbeault cohomology group
$H^{n,n-2}_{\bar\partial_A ;P}(Z)$.  Thus, non-Abelian holomorphic BF
theories give a field-theoretic description of holomorphic structures
on  bundles $P\to Z$ and of the twisted Dolbeault complex on $Z$.

\subsection{Abelian models}

In the Abelian case, instead of a $\fg$-valued $(0,1)$-connection on
a bundle over a complex manifold $Z$ one can take any $(p,q-1)$-form
$A^{p,q-1}\in\Omega^{p,q-1}(Z)$ and introduce the action
$$
S_{\mathrm{hAB}}=\int_ZB^{n-p,n-q}\wedge\bar\partial A^{p,q-1},
\eqno(2.9)
$$
where $B^{n-p,n-q}\in \Omega^{n-p,n-q}(Z)$, i.e. $B^{n-p,n-q}$ is a
$\C$-valued $(n-p,n-q)$-form on $Z$. Equations of motion following
from (2.9) are
$$
\bar\partial A^{p,q-1}=0,
\eqno(2.10a)
$$
$$
\bar\partial B^{n-p,n-q}=0.
\eqno(2.10b)
$$

The action (2.9) and Eqs.(2.10) are invariant under the following
symmetry transformations:  $$ A^{p,q-1}\mapsto A^{p,q-1}+\bar\partial
\Phi^{p,q-2}, \eqno(2.11a) $$ $$ B^{n-p,n-q}\mapsto
B^{n-p,n-q}+\bar\partial \Phi^{n-p,n-q-1}, \eqno(2.11b) $$ where
$\Phi^{r,s}\in \Omega^{r,s}(Z)$. Therefore the moduli space of
solutions to Eqs.(2.10) is a vector space $$
\cm_{\mathrm{hAB}}=H^{p,q-1}_{\bar\partial}(Z)\oplus
H^{n-p,n-q}_{\bar\partial}(Z), \eqno(2.12) $$ where
$H^{r,s}_{\bar\partial}(Z)$ is the $(r,s)$th Dolbeault cohomology
group of $Z$. So, the action (2.9) provides us with a
field-theoretic description of the standard Dolbeault complex on $Z$.

\section{Special holomorphic BF theories}
\subsection{Hermitian-Yang-Mills connections and stable bundles}

Let $Z$ be a K\"ahler $n$-manifold with a K\"ahler form $\o$,
and $G_\R$ be a subgroup of the unitary group $U(l)$ such that
$(G_\R)^\C =G_\R\otimes\C=G, l\ge 2$. We consider a (trivial)
principal $G_\R$-bundle
$P_\R$ over $Z$ and the $(0,1)$-component $A^{0,1}$ of a unitary
connection 1-form $A=A^{1,0}+ A^{0,1}$ on $P_\R$.

\smallskip

Recall that any unitary connection $A$ on $P_\R$ defines a
holomorphic structure on a $G$-bundle $P$ over $Z$ if the
$(0,1)$-component $A^{0,1}$ of $A$ is flat, i.e. satisfies
Eqs.(2.2a). Connections $A$ are called {\it Hermitian-Yang-Mills
connections}, if their curvature $F_A$ satisfies the following
equations~\cite{Do}:
$$
F^{0,2}_A=0,
\eqno(3.1a)
$$
$$
\o^{n-1}\wedge F^{1,1}_A=0,
\eqno(3.1b)
$$
where
$$
F^{0,2}_A=\bar\p A^{0,1} + A^{0,1}\wedge A^{0,1},
\eqno(3.2a)
$$
$$
F^{1,1}_A=\p A^{0,1} +\bar\p A^{1,0}+ A^{1,0}\wedge A^{0,1}+
A^{0,1}\wedge A^{1,0}.
\eqno(3.2b)
$$
Hermitian-Yang-Mills connections correspond to {\it stable} holomorphic
structures  $\bar\p_A$ on a bundle~\cite{Do}, but this is in some
sense most of them.

\smallskip

Notice that Eqs.(3.1) can be rewritten in the form
$$
F^{0,2}_A=0,
\eqno(3.3a)
$$
$$
\Lambda F^{1,1}_A=0,
\eqno(3.3b)
$$
where $\Lambda$ is an algebraic `trace' operator which measures the
component of a $(1,1)$-form parallel to $\o$~\cite{Do}.
In local coordinates $\{z^a\}$ on $Z$, Eqs.(3.3) have the form
$$F_{\bar a\bar b}=0,
\eqno(3.4a)
$$
$$
g^{a\bar b}F_{a\bar b}=0,
\eqno(3.4b)
$$
where ${\mathbf{g}}=\{g^{a\bar b}\}$ is a Riemannian metric
compatible with the complex structure on $Z$. Equations (3.3) appear
in superstring theory from the condition of preserving at least one
unbroken supersymmetry in four dimensions after compactification of
$D=10$ superstrings on Calabi-Yau 3-folds~\cite{GSW}.

\smallskip

The study of Eqs.(3.1) or equivalent Eqs.(3.3) is of interest because
of their importance in algebraic geometry and superstring theory. We
introduce an action that leads to equations of motion containing
Eqs.(3.1). Namely, consider the action $$
S_{\mathrm{HYM}}=S_{\mathrm{hBF}}+\b S_{\o\Xi\mathrm{F}} =\int_{Z}\Tr
(B\wedge F^{0,2}_A-\b\Xi\o^{n-1} \wedge F^{1,1}_A), \eqno(3.5) $$
where $\Xi$ is a smooth $\fg$-valued function on $Z$ and $\b$ is a
constant parameter. The action (3.5) is a one-parameter deformation
of the action $S_{\mathrm{hBF}}$ for holomorphic BF theories
introduced in Sect.2.1. Let us emphasize that the action
$S_{\mathrm{HYM}}$ depends on a K\"ahler form $\o$ on $Z$.

\smallskip

The field equations following from the action (3.5) are
$$F^{0,2}_A=0,  \quad
\b\o^{n-1}\wedge F^{1,1}_A=0,
\eqno(3.6a)
$$
$$
\bar\p B+ A^{0,1}\wedge B - B\wedge A^{0,1}= \p (\b\Xi\o^{n-1})+
\o^{n-1}\wedge [A^{1,0}, \b\Xi ],
\eqno(3.6b)
$$
$$
\bar\p (\b\Xi\o^{n-1})+      \o^{n-1}\wedge [A^{0,1}, \b\Xi ]=0,
\eqno(3.6c)
$$
where $[,]$ is the standard commutator in the Lie algebra $\fg$.
We see that Eqs.(3.6a) coincide with Eqs.(3.1) and therefore
$A=A^{1,0}+A^{0,1}$ is a Hermitian-Yang-Mills connection. It is
obvious that if $\b =0$, then Eqs.(3.6) coincide with Eqs.(2.2)
of a holomorphic BF theory.

\subsection{Holomorphic $\th$BF theories on Calabi-Yau manifolds}

Let now $Z$ be a Calabi-Yau $n$-fold. This means that besides a
complex structure, on $Z$ there exist a K\"ahler 2-form $\o$, a
Ricci-flat K\"ahler metric $\mathbf g$ and a nowhere vanishing
holomorphic $(n,0)$-form $\th$.  The $(n,0)$-form $\th$ can be taken
to be covariantly constant w.r.t. the Levi-Civita connection. We
consider a (trivial) principal $G$-bundle $P$ over $Z$ and the
$(0,1)$-component $A^{0,1}$ of a connection 1-form $A$ on $P$.

\smallskip

{}For bundles $P$ over Calabi-Yau (CY) $n$-folds $Z$ one can consider
holomorphic BF theories with the action (2.1) which does not depend
on K\"ahler structures, metrics and holomorphic $(n,0)$-forms on $Z$.
But the existence on $Z$ of a nowhere degenerate holomorphic
$(n,0)$-form $\th$ permits one to introduce one more class of models
which will be called {\it holomorphic $\th$BF theories}. Their action
has the form $$ S_{\mathrm{h}\th\mathrm{BF}}=\int_Z\th\wedge
\Tr(B\wedge F^{0,2}_A), \eqno(3.7) $$ where $B$ is a $\fg$-valued
$(0,n-2)$-form on $Z$, and $F^{0,2}_A= \bar\p A^{0,1}+ A^{0,1}\wedge
A^{0,1}$ is the $(0,2)$-component of the curvature tensor of a
connection $A$ on $P$. The action (3.7) leads to the equations of
motion $$ \th\wedge F^{0,2}_A=0,\quad \th\wedge (\bar\p B+
A^{0,1}\wedge B - (-1)^nB\wedge A^{0,1})=0 \eqno(3.8) $$ which are
equivalent to the equations $$ \bar\p A^{0,1}+A^{0,1}\wedge
A^{0,1}=0, \eqno(3.9a) $$ $$ \bar\p B+ A^{0,1}\wedge B -
(-1)^nB\wedge A^{0,1}=0.  \eqno(3.9b) $$ In abridged notation,
Eqs.(3.9b) can be rewritten as $\bar\p_A B=0$.

\smallskip

{}From Eqs.(3.9) we see that holomorphic $\th$BF theories describe
holomorphic structures on bundles $P\to Z$ and $\bar\p_A$-closed forms
$B\in\Omega^{0,n-2}(Z,\fg )$. Solutions to Eqs.(3.9a)
are flat (0,1)-connections and for any fixed flat $(0,1)$-connection
$A^{0,1}$ on $P$ the space of nontrivial solutions
to Eqs.(3.9b) is the $(0,n-2)$th Dolbeault cohomology group
$$
H^{0,n-2}_{\bar\p_A; P}:=\frac{\{B\in\Omega^{0,n-2}(Z,\fg ):
\bar\p_A B=0\}} {\{B=\bar\p_A \Phi , \Phi\in \Omega^{0,n-3}(Z,\fg )\}}.
\eqno(3.10)
$$
So, the action (3.7) provides us with a field-theoretic description
of the Dolbeault complex coupled to a flat $(0,1)$-connection on the
adjoint bundle $\Ad P$ over a Calabi-Yau $n$-fold $Z$.

\subsection{Special hBF theories on twistor spaces of self-dual
4-manifolds}

Let us consider a Riemannian real 4-manifold $M$ with self-dual Weyl
tensor (a {\it self-dual manifold}) and the bundle $\tau : Z\to M$ of
complex structures on $M$ (the {\it twistor space} of $M$) with $\C
P^1$ as a typical fiber~\cite{Pe, AHS}. The twistor space $Z$ of a
self-dual 4-manifold $M$ is a complex 3-manifold~\cite{AHS} which is
the total space of a fiber bundle over $M$ associated with the bundle
of orthonormal frames on $M$.

\smallskip

It is well known that Yang-Mills instantons on $M$ can be described
in terms of holomorphic bundles over the twistor space $Z$ of $M$.
Namely, there is a one-to-one correspondence between {\it self-dual
bundles} (bundles with self-dual connections) over $M$ and
holomorphic bundles $P$ over $Z$ that are holomorphically trivial on
any projective line $\C P^1_x\hra Z$ parametrized by $x\in
M$~\cite{Wa1,AW,AHS}.  The condition of holomorphic
trivializability of the bundle $P$ after the restriction of $P$ to
every projective line in $Z$ is equivalent to the equality to zero of
the restriction of the $(0,1)$-component $A^{0,1}$ of a connection
$A$ on $P$ to submanifolds $\C P^1_x\hra Z, x\in M$ (see~\cite{Po1}
for discussions). Such connections on $P$ can be described in terms
of holomorphic BF theories on $Z$ by adding a term to the action
(2.1). In this way we obtain a model which describes the instanton
moduli space and differs from the Donaldson-Witten model.

\smallskip

Consider the twistor space $Z$ of a self-dual real 4-manifold $M$. By
definition, $Z$ is a complex fibered 3-manifold with the canonical
projection $\tau : Z\to M$. The typical fiber $\C P^1$ has the
$SU(2)$-invariant complex structure (see~\cite{Po2} for more
details), and the vertical distribution $\cv=\Ker\tau_*$ inherits
this complex structure. A restriction of $\cv$ to each fiber $\C
P^1_x$, $x\in M$, is the tangent bundle to that fiber. The
Levi-Civita connection on the Riemannian manifold $M$ generates the
splitting of the tangent bundle $T(Z)$ into a direct sum $$
T(Z)=\cv\oplus\ch \eqno(3.11) $$ of the vertical $\cv$ and horizontal
$\ch$ distributions. Using the complex structures on $\C P^1$ and
$Z$, one can split the complexified tangent bundle of $Z$ into a
direct sum $$ T^\C(Z)=T^{1,0}\oplus
T^{0,1}=(\cv^{1,0}\oplus\ch^{1,0})\oplus (\cv^{0,1}\oplus\ch^{0,1})
\eqno(3.12) $$ of subbundles of type $(1,0)$ and $(0,1)$. Analogously
one can split the complexified cotangent bundle of $Z$ into a direct
sum of subbundles $T_{1,0}$ and $T_{0,1}$. So we have the
(integrable) distribution $\cv^{0,1}$ of $(0,1)$-vector fields.

\smallskip

Let $E^{3,3}$ be a $\fg$-valued $(3,3)$-form on $Z$, and $V^{0,1}$ be
an arbitrary (0,1)-vector field from the distribution
$\cv^{0,1}$. Denote by $\{e_a\}$, $\{e_{\bar a}\}$, $\{\s^a\}$ and
$\{\s^{\bar a}\}$ local frames for the bundles $T^{1,0}$, $T^{0,1}$,
$T_{1,0}$ and $T_{0,1}$, respectively. Then locally
$$E^{3,3}=E_{a_1...\bar a_3}\s^{a_1}\wedge ...\wedge\s^{\bar
a_3},\quad V^{0,1}=V^{\bar a}e_{\bar a},\quad A^{0,1}=A^{0,1}_{\bar a}
\sigma^{\bar a}, \eqno(3.13) $$ where $a_1,\bar a_1,...=1,2,3$.
Let us consider the contraction $$
V^{0,1}\contra E^{3,3} \eqno(3.14a) $$ of $V^{0,1}$ with $E^{3,3}$,
which is a $\fg$-valued $(3,2)$-form on $Z$.  In the local frames,
the form (3.14a) has components $$ (V^{0,1}\contra
E^{3,3})_{a_1a_2a_3\bar a_1\bar a_2}:= 3V^{\bar a_3} E_{a_1a_2a_3\bar
a_1\bar a_2\bar a_3} .  \eqno(3.14b) $$ We also introduce a
$\fg$-valued function $$ V^{0,1}\contra A^{0,1}=V^{\bar a}A^{0,1}_{\bar a}
\eqno(3.14c) $$ on $Z$ and notice that $$ (V^{0,1}\contra
E^{3,3})\wedge A^{0,1}=-E^{3,3} (V^{0,1}\contra A^{0,1}) \eqno(3.15)
$$ since $E^{3,3}\wedge A^{0,1}\equiv 0$.

\smallskip

To the action (2.1) of a holomorphic BF theory we add the term $$
S_{\mathrm{VAE}}=\g\int_Z\Tr[(V^{0,1}\contra E^{3,3})\wedge A^{0,1}],
\eqno(3.16) $$ where $\g =const$.  The action $$
S_{\mathrm{hBF}}+S_{\mathrm{VAE}}= \int_Z\Tr[B^{3,1}\wedge
F^{0,2}_A-\g (V^{0,1}\contra A^{0,1})E^{3,3}] \eqno(3.17) $$ leads to
the following field equations:  $$ \bar\p A^{0,1}+A^{0,1}\wedge
A^{0,1}=0,\quad \g  V^{0,1}\contra A^{0,1}=0, \eqno(3.18a) $$ $$
\bar\p B^{3,1}+A^{0,1}\wedge B^{3,1} - B^{3,1}\wedge A^{0,1}= \g
V^{0,1}\contra E^{3,3}.  \eqno(3.18b) $$ Equations (3.18a) on the
twistor space $Z$ of a self-dual 4-manifold $M$ are equivalent to the
self-dual Yang-Mills (SDYM) equations on $M$, which was discussed in
detail in~\cite{Po1,Po2}. Equations (3.18b) describe some fields
interacting with external self-dual gauge fields on $M$.  It is
obvious that in the limit $\g\to 0$, Eqs.(3.18) are reduced to
Eqs.(2.2) of holomorphic BF theories on $Z$.

\subsection{Conformal and integrable field theories related to special
hBF theories}

It is well known that Chern-Simons theories on $3D$ {\it real}
manifolds $X$ are reduced to $2D$ conformal field theories (CFT) if
one supposed that a 3-manifold $X$ has the form of a trivial {\it
bundle} $\Sigma\times\R$ (or $\Sigma\times S^1$), where $\Sigma$ is a
2-manifold~\cite{W2,MoS}.  It is reasonable to expect that
holomorphic analogs of Chern-Simons and BF theories defined on a $3D$
{\it complex} manifold $Z$ (six real dimensions) can be reduced to
$4D$ CFT's if one supposes that $Z$ is a bundle over a real
4-manifold $M$ with two-dimensional fibers. One important example of
a conformal field theory in four dimensions is provided by the SDYM
model. In Sect.3.3 we have described the reduction of holomorphic BF
models on $Z$ to SDYM models on self-dual 4-manifolds $M$. For
discussions of reductions to other $4D$ CFT's, see~\cite{Po1}.

\smallskip

Notice that Hermitian-Yang-Mills connections (see Sect.3.1) on
bundles $P$ over complex 2-dimensional K\"ahler manifolds are
self-dual connections since Eqs.(3.6a)  for $n=2$ coincide with the
SDYM equations.  Thus, both the action (3.5) with $n=2$ and the
action (3.17) describe SDYM models on 4-manifolds. In the simplest
case of the flat 4-space, we obtain SDYM models on $\R^4$. It is well
known that almost all integrable equations in $1\le D\le 3$
dimensions (Bogomolny, Korteweg-de Vries, Nonlinear Schr\"odinger,
Boussinesq, Nahm and many others) can be obtained by reductions of 
the SDYM equations
on $\R^4$ (see~\cite{Wa2, MW} and references therein).
Thus, special holomorphic BF theories are connected with $4D$ CFT's
and integrable models in $1\le D\le 4$ dimensions.

\newpage
\section{Observables and topological invariants}

To introduce observables and topological invariants for holomorphic 
BF and $\th$BF theories we use the
results on holomorphic analogs of Chern-Simons theories defined on
Calabi-Yau 3-folds~\cite{W3, BCOV, DT, Th}.

\smallskip

{}First, for an action $S$ coinciding with (2.1), (2.9) or (3.7), one
can write down the partition function $$ Z_k=\int_{\ca /\cg}
\cd A^{0,1}\exp (ikS), \eqno(4.1) $$ where $k\in\Z$, and the path
integral is evaluated over the space $\ca /\cg$ of gauge inequivalent
$(0,1)$-connections $A^{0,1}$ on a bundle $P\to Z$.

\smallskip

Second, using the semiclassical approximation, one can give the 
partition function interpretation of the Ray-Singer holomorphic 
torsion~\cite{RS} of the holomorphic bundle $(\Ad P, \bar\p_A)$.

\smallskip

Third, one can introduce an analog of Wilson loops for Abelian
$(0,1)$-connections  $A^{0,1}$ on a line bundle $L\to Z$~\cite{Th}.
For this one should fix complex curves $C_i\subset Z$ (e.g. tori),
holomorphic 1-forms $\a_i$ on the $C_i$'s and define
$$
T(C_i)=\exp (ik\int_{C_i} \a_i\wedge A^{0,1})
\eqno(4.2)
$$
as an element of the Jacobian of $L$ paired against $\a_i$
and exponentiated. Then we can consider the path integral
$$
\int_{\ca /\cg} \cd A^{0,1}\exp (ikS)\mathop{\prod}^m_{i=1}T(C_i)
\eqno(4.3)
$$
and try to prove that it is an invariant of the complex structure of a
manifold $Z$.

\smallskip

{}Finally, for Abelian holomorphic BF theories with the action (2.9)
one can introduce the Dolbeault currents~\cite{GrH} and define the
linking number of complex submanifolds of a manifold $Z$ using a pairing
$H^{p,q}_{\bar\p}(Z)\otimes H^{n-p,n-q}_{\bar\p}(Z) \to \C$ of the
Dolbeault cohomology groups.

\smallskip

\section{Conclusion}

In this paper we have introduced new classes of gauge field theories
which are natural holomorphic analogs of BF topological theories.
These holomorphic BF theories give a field-theoretic description of
holomorphic structures on bundles over complex $n$-manifolds. Three
special holomorphic BF theories on K\"ahler, Calabi-Yau and fibered
complex manifolds have been introduced.  In particular, we have
introduced an action describing Hermitian-Yang-Mills connections on
stable holomorphic bundles.  Such connections satisfy the
Donaldson-Uhlenbeck-Yau equations and are often used in superstring
theory in the description of compactified configurations with
unbroken supersymmetry in four dimensions.

\smallskip

There are many open problems that have to be considered. In
particular, it is necessary to analyze the ghost structure of the
introduced theories, to perform BRST gauge fixing and write down
quantum actions.  Supersymmetrization of holomorphic BF theories may
be of interest.  The problem of finding nontrivial observables and
their metric independence should be more carefully analyzed. Much
work remains to be done.

\end{document}